\documentclass[aps,pre,reprint,groupedaddress]{revtex4-1}

\usepackage{amsmath,amssymb}
\usepackage{bm}
\usepackage{xcolor}
\usepackage{graphicx}
\usepackage[dvipdfmx]{hyperref}
\hypersetup{colorlinks=true,citecolor=magenta,linkcolor=magenta,urlcolor=magenta}

\begin{document}


\title{Iterative classical superadiabatic algorithm for combinatorial optimization}


\author{Takuya Hatomura}
\email[]{takuya.hatomura.ub@hco.ntt.co.jp}
\affiliation{NTT Basic Research Laboratories, NTT Corporation, Kanagawa 243-0198, Japan}


\date{\today}

\begin{abstract}
We consider a classical and superadiabatic version of an iterative quantum adiabatic algorithm to solve combinatorial optimization problems. 
This algorithm is deterministic because it is based on purely classical dynamics, that is, it does not rely on any stochastic approach to mimic quantum dynamics. 
Moreover, use of shortcuts to adiabaticity makes the algorithm independent of the annealing time. 
We apply this algorithm to a certain class of hard instances of the 3-SAT problem. 
We find that more than 90\% of such 64-bits hard instances can be resolved by a few iteration. 
Our approach can also be used to analyze properties of instances themselves apart from stochastic uncertainty and shortage of adiabaticity. 
\end{abstract}


\maketitle
\onecolumngrid
%
%
\section{Introduction}
A quantum adiabatic algorithm, which is a prototype of adiabatic quantum computation, was proposed as a method to solve combinatorial optimization problems based on quantum mechanics~\cite{Farhi2000,Farhi2001}. 
The solution of a given problem is encoded in the ground state of a problem Hamiltonian. 
Then, the trivial ground state of a driver Hamiltonian is adiabatically transformed into the ground state of the problem Hamiltonian. 
This algorithm is also known as quantum annealing~\cite{Kadowaki1998}, where a transverse field was introduced as a driver Hamiltonian to find the ground state of Ising spin glasses. 
Because many combinatorial optimization problems can be formulated as a problem finding the ground state of Ising spin glasses~\cite{Lucas2014}, these two algorithms are in principle the same, while these two algorithms have been developed toward different directions.

Performance of the quantum adiabatic algorithm depends on energy difference between the ground state and higher energy eigenstates of the total Hamiltonian because of the adiabatic theorem~\cite{Kato1950}. 
Therefore, existence of a quantum phase transition during adiabatic transformation, which causes a small energy gap, is a bottleneck of this algorithm. 
For such problems showing quantum phase transitions, we can use different paths~\cite{Farhi2002}, for example, different choice of a driver Hamiltonian or a different way of intervention, to mitigate hardness. 
One of the easiest ways constructing different paths and performing annealing processes following those paths is to apply a random transverse field as a driver Hamiltonian~\cite{Farhi2011}, which is recently known as inhomogeneous driving.

By using numerical and perturbative analysis, removal of a small energy gap associated with an energy crossing was found in the 3-SAT problem, where a random transverse field was used~\cite{Farhi2011}. 
One might be interested in if all small energy gaps can be removed by using a random transverse field or not. 
For the maximum independent set problem, existence of a path that has no energy crossing during adiabatic transformation was proved by using perturbative expansion~\cite{Dickson2011}. 
Here, this path is modified by an inhomogeneous transverse field. 
Heuristic scheduling of inhomogeneous driving realizing such a path is possible by using an iterative quantum algorithm~\cite{Dickson2012}. 
In addition, mitigation of a small energy gap by using inhomogeneous driving was experimentally demonstrated~\cite{Lanting2017}. 
It is also possible to design inhomogeneous driving by using the knowledge of interactions in a problem Hamiltonian~\cite{Adame2018}. 
Furthermore, inhomogeneous driving has also been studied from other theoretical viewpoints. 
Effect of inhomogeneity on topological defects and how to suppress those defects were studied from the viewpoint of the inhomogeneous Kibble-Zurek mechanism~\cite{Rams2016a,Mohseni2018}. 
Moreover, by using inhomogeneous driving, removal of first order transitions during adiabatic transformation, which cause exponentially small energy gaps, was analytically shown in the $p$-spin model~\cite{Susa2018,Susa2018a} and the Lechner-Hauke-Zoller model~\cite{Hartmann2019}. 
In addition, removal of energy crossings in a classical model of quantum annealing, where the two-dimensional random field Ising model was considered as a problem Hamiltonian, was also found, while to find such a path is exponentially hard~\cite{Hatomura2018b}.

In this paper, we consider a certain class of hard instances of the 3-SAT problem~\cite{Farhi2011}, which show small energy gaps associated with energy crossings during adiabatic transformation. 
By using shortcuts to adiabaticity~\cite{Guery-Odelin2019} for classical spin systems, of which the present author and a coworker found the exact construction~\cite{Hatomura2018b}, we confirm that energy crossings also arise in a classical model of the quantum adiabatic algorithm (quantum annealing), and thus those instances are also classically hard. 
By using our algorithm, we also check when such energy crossings in the classical model happen during adiabatic transformation. 
Moreover, to remove those energy crossings, we consider inhomogeneous driving, where we update a transverse field according to a certain rule which is similar to that of an iterative quantum adiabatic algorithm~\cite{Dickson2012}. 
We show that large amount of those hard instances can be resolved by a few iteration even in the classical model. 
Note that our algorithm is deterministic because we do not use any stochastic approach to simulate quantum dynamics, such as quantum Monte Carlo simulation~\cite{Santoro2002,Martonak2002}, and it is independent of the annealing time because we use exact construction of shortcuts to adiabaticity (exact construction of shortcuts to adiabaticity for many-body systems is limited to a few models, see, e.g., Ref.~\cite{DelCampo2012}), and thus we can study properties of a given problem without stochastic uncertainty and shortage of adiabaticity.

The rest of the present article is constructed as follows. 
In Sec.~\ref{Sec.QAA}, we review the quantum adiabatic algorithm (quantum annealing) and introduce the 3-SAT problem. 
A method to generate a certain class of hard instances is also explained. 
In Sec.~\ref{Sec.CSA}, we introduce a classical version of the quantum adiabatic algorithm and application of shortcuts to adiabaticity for classical spin systems to it. 
We will see how to solve the equation of motion for counterdiabatic dynamics. 
We briefly review an iterative quantum adiabatic algorithm in Sec.~\ref{Sec.IQAA} and give a similar way of iteration for the classical superadiabatic algorithm in Sec.~\ref{Sec.ite.classical}. 
We perform numerical simulation in Sec.~\ref{Sec.results}. 
First, we check if hard instances are successfully generated or not in Sec.~\ref{Sec.planted}. 
We also confirm when energy crossings happen during adiabatic transformation. 
Next, we try to resolve those hard instances by iteration in Sec.~\ref{Sec.resolve.hard}. 
We summarize and discuss our results in Sec.~\ref{Sec.discuss}. 

%
%
\section{(Super)Adiabatic algorithms}
%
%
\subsection{\label{Sec.QAA}Quantum adiabatic algorithm}
In a quantum adiabatic algorithm~\cite{Farhi2000,Farhi2001} and quantum annealing~\cite{Kadowaki1998}, we encode the solution of a given problem in the ground state of a problem Hamiltonian $\hat{\mathcal{H}}_P$ and prepare a driver Hamiltonian $\hat{\mathcal{H}}_V$ whose ground state is trivial. 
These two Hamiltonians are intervened by a continuous function $g(t/T)$ as
\begin{equation}
\hat{\mathcal{H}}(t)=g(t/T)\hat{\mathcal{H}}_P+[1-g(t/T)]\hat{\mathcal{H}}_V,
\label{Eq.QAham}
\end{equation}
where $g(t/T)$ satisfies $g(0)=0$ and $g(1)=1$ so that $\hat{\mathcal{H}}(0)=\hat{\mathcal{H}}_V$ and $\hat{\mathcal{H}}(T)=\hat{\mathcal{H}}_P$. 
Here, $T$ is the annealing time, which is the runtime of this algorithm. 
The adiabatic theorem~\cite{Kato1950} indicates that starting from the trivial ground state of the driver Hamiltonian we can obtain the ground state of the problem Hamiltonian, that is, the solution of a given problem if the annealing time $T$ is large enough. 
A drawback of this algorithm is presence of small energy gaps during adiabatic transformation, which requires exponentially large annealing time. 
Note that use of different paths, for example, different choice of a driver Hamiltonian or a different way of intervention, can sometimes resolve this hardness~\cite{Farhi2002,Farhi2011,Dickson2011,Dickson2012,Lanting2017,Adame2018,Rams2016a,Mohseni2018,Susa2018,Susa2018a,Hartmann2019,Hatomura2018b}.

Following Ref.~\cite{Farhi2011}, we consider the 3-SAT problem. 
The SAT problem asks if there is any bit string that satisfies given clauses or not. 
If each clause consists of three literals at most, this problem is called the 3-SAT problem. 
Here we consider $N$-bit strings $b_1b_2\cdots b_N$, where $b_i=0$ or $1$, and $M$ clauses $C_1\land C_2\land\cdots\land C_M$. 
Suppose that each clause $C_m$ imposes a penalty $1$ on a set of three bits $i_m,j_m,k_m\in\{1,2,\cdots,N\}$ if its bit string $b_{i_m}b_{j_m}b_{k_m}$ is identical to a given value $w_{i_m}w_{j_m}w_{k_m}$, where $w_i=0$ or $1$, that is, it can be written as $C_m=(\lnot(b_{i_m}=w_{i_m}))\lor(\lnot(b_{j_m}=w_{j_m}))\lor(\lnot(b_{k_m}=w_{k_m}))$. 
The problem Hamiltonian of this problem is expressed as
\begin{equation}
\hat{\mathcal{H}}_P=\sum_{m=1}^M\frac{1+(-1)^{w_{i_m}}\hat{\sigma}_{i_m}^z}{2}\frac{1+(-1)^{w_{j_m}}\hat{\sigma}_{j_m}^z}{2}\frac{1+(-1)^{w_{k_m}}\hat{\sigma}_{k_m}^z}{2},
\end{equation}
where $\hat{\sigma}_i^\alpha$, $\alpha=x,y,z$, is the Pauli matrix. 
Here a bit $b_i=0$ ($b_i=1$) corresponds to a state $|\uparrow\rangle_i$ ($|\downarrow\rangle_i$), where $\hat{\sigma}_i^z|\uparrow\rangle_i=+|\uparrow\rangle_i$ ($\hat{\sigma}_i^z|\downarrow\rangle_i=-|\downarrow\rangle_i$). 
Indeed, each term gives $1$ if $b_{i}b_{j}b_{k}=w_{i}w_{j}w_{k}$, and otherwise it gives $0$. 
For a driver Hamiltonian, we use the transverse field Hamiltonian
\begin{equation}
\hat{\mathcal{H}}_V=\sum_{i=1}^N\Delta_i\frac{1-\hat{\sigma}_i^x}{2}, 
\end{equation}
where a set of positive constants $\{\Delta_i\}$ is tuned to avoid small energy gaps in inhomogeneous driving.

A way to generate hard instances was also demonstrated in Ref.~\cite{Farhi2011}. 
We first randomly produce clauses penalizing the following values $w_{i}w_{j}w_{k}\in\{100,010,001,110,101,011\}$ with enough large number of clauses. 
Then, two bit strings $00\cdots0$ and $11\cdots1$ become the degenerate ground states. 
According to Ref.~\cite{Farhi2011}, number of clauses should be $M\sim5N\log N$ to exclude other satisfying assignments. 
We apply the second order perturbation theory for a small uniform transverse field and determine the lower energy state among two degenerate ground states. 
Then, we add a small penalty to the lower energy state such as $[(1+\hat{\sigma}_1^z)/2][(1+\hat{\sigma}_2^z)/2][(1+\hat{\sigma}_3^z)/2]/2$ for $00\cdots0$ or $[(1-\hat{\sigma}_1^z)/2][(1-\hat{\sigma}_2^z)/2][(1-\hat{\sigma}_3^z)/2]/2$ for $11\cdots1$. 
This small penalty lifts degeneracy, reverses the lowest and the second lowest levels, and produces a small energy gap associated with the energy crossing between the bit strings $00\cdots0$ and $11\cdots1$. 
By using numerical and perturbative analysis, it was shown that random choice of inhomogeneity $\{\Delta_i\}$ resolves this small energy gap~\cite{Farhi2011}. 
Later, we will show that this generation method also creates an energy crossing in a classical model of quantum adiabatic algorithm and it can be removed by using inhomogeneous driving even in a classical framework.

%
%
\subsection{\label{Sec.CSA}Classical superadiabatic algorithm}
First, we introduce shortcuts to adiabaticity for classical spin systems, of which the exact construction was found by the present author and a coworker~\cite{Hatomura2018b}. 
We consider a classical spin system consisting of $N$ classical spins expressed by three dimensional unit vectors $\{\bm{m}_i\}_{i=1}^N$, where $|\bm{m}_i|=1$. 
Suppose that dynamics of this system is governed by a time-dependent classical Hamiltonian $\mathcal{H}_t(\{\bm{m}_i\})$. 
For this Hamiltonian, the counterdiabatic Hamiltonian, which counteracts diabatic changes~\cite{Demirplak2003,Berry2009}, is given by
\begin{equation}
\mathcal{H}_t^\mathrm{cd}(\{\bm{m}_i\})=\sum_{i=1}^N\bm{f}_i(t)\cdot\bm{m}_i,
\label{Eq.cdHam}
\end{equation}
where
\begin{equation}
\bm{f}_i(t)=\frac{\bm{h}_i^\mathrm{eff}(t)\times\dot{\bm{h}}_i^\mathrm{eff}(t)}{2|\bm{h}_i^\mathrm{eff}(t)|^2}, 
\end{equation}
(for derivation, see Ref.~\cite{Hatomura2018b}). 
Here, $\bm{h}_i^\mathrm{eff}(t)$ is the effective field
\begin{equation}
\bm{h}_i^\mathrm{eff}(t)=-\frac{\partial\mathcal{H}_t}{\partial\bm{m}_i}. 
\end{equation}
Then, under the total Hamiltonian $\mathcal{H}_t^\mathrm{tot}(\{\bm{m}_i\})=\mathcal{H}_t(\{\bm{m}_i\})+\mathcal{H}_t^\mathrm{cd}(\{\bm{m}_i\})$, the equation of motion is given by the torque equation
\begin{equation}
\dot{\bm{m}}_i(t)=2\bm{m}_i(t)\times[\bm{h}_i^\mathrm{eff}(t)-\bm{f}_i(t)]. 
\label{Eq.EOM}
\end{equation}
It was shown that the solution of the equation of motion conserves angles between each classical spin and its effective field irrespective of time dependence of the original Hamiltonian $\mathcal{H}_t(\{\bm{m}_i\})$ (for proof, see Ref.~\cite{Hatomura2018b}). 
Therefore, starting from a stationary state, where angles between each classical spin and its effective field are zero, the system always follows an instantaneous stationary state as long as it exists.

Now we introduce a classical model of the quantum adiabatic algorithm~\cite{Hatomura2018b}. 
We define the classical Hamiltonian $\mathcal{H}_t(\{\bm{m}_i\})$ by replacing the Pauli matrices $\{\bm{\sigma}_i\}_{i=1}^N$ in the quantum Hamiltonian of the quantum adiabatic algorithm (\ref{Eq.QAham}) with these classical spins $\{\bm{m}_i\}_{i=1}^N$, that is, the equality $\hat{\mathcal{H}}(t)=\mathcal{H}_t(\{\hat{\bm{\sigma}}_i\})$ holds. 
Here we assume that each spin does not interact with itself. 
This classical Hamiltonian $\mathcal{H}_t(\{\bm{m}_i\})=g(t/T)\mathcal{H}_P+[1-g(t/T)]\mathcal{H}_V$ is actually the classical limit of the quantum Hamiltonian (\ref{Eq.QAham}) because the canonical equation of motion for this classical Hamiltonian coincides with the classical limit of the Heisenberg equation of motion under the quantum Hamiltonian. 
Note that the ground state of the classical problem Hamiltonian $\mathcal{H}_P$ is identical to that of the quantum problem Hamiltonian $\hat{\mathcal{H}}_P$. 
An exception is that the quantum problem Hamiltonian can have a superposition of degenerate ground states. 
However, even in that case, each degenerate ground state expressed in the computational basis is a degenerate ground state of the classical problem Hamiltonian. 
It should be also noted that there is no guarantee to find the ground state of the problem Hamiltonian by using this classical model of the quantum adiabatic algorithm even if the annealing time is infinitely large or even if the above shortcut method is used. 
This is because the ground state becomes a metastable state when an energy crossing happens.

Here we explain in detail how to solve this equation of motion (\ref{Eq.EOM}) under the classical Hamiltonian $\mathcal{H}_t(\{\bm{m}_i\})$ corresponding to the quantum Hamiltonian (\ref{Eq.QAham}). 
Because the Hamiltonian $\mathcal{H}_t(\{\bm{m}_i\})$ does not include $m_i^y$, the effective field $\bm{h}_i^\mathrm{eff}(t)$ does not have the $y$ component and the counterdiabatic field $\bm{f}_i(t)$ only has the $y$ component. 
Therefore, the equation of motion (\ref{Eq.EOM}) is rewritten as
\begin{equation}
\left\{
\begin{aligned}
&\dot{m}_i^x(t)=2m_i^y(t)h_i^{\mathrm{eff},z}(t)+2m_i^z(t)f_i^y(t), \\
&\dot{m}_i^y(t)=2m_i^z(t)h_i^{\mathrm{eff},x}(t)-2m_i^x(t)h_i^{\mathrm{eff},z}(t), \\
&\dot{m}_i^z(t)=-2m_i^x(t)f_i^y(t)-2m_i^y(t)h_i^{\mathrm{eff},x}(t), 
\end{aligned}
\right.
\label{Eq.EOM.detail}
\end{equation}
where $f_i^y(t)$ is given by
\begin{equation}
f_i^y(t)=\frac{h_i^{\mathrm{eff},z}(t)\dot{h}_i^{\mathrm{eff},x}(t)-h_i^{\mathrm{eff},x}(t)\dot{h}_i^{\mathrm{eff},z}(t)}{2|\bm{h}_i^\mathrm{eff}(t)|^2}. 
\end{equation}
Because the $z$ component of the time derivative of the effective field $\dot{h}_i^{\mathrm{eff},z}(t)$ includes $\{\dot{m}_j^z\}$, first we have to solve the simultaneous equations of motion (\ref{Eq.EOM.detail}) with respect to $\{\dot{\bm{m}}_i\}$. 
The $z$ component of the time derivative of the effective field $\dot{h}_i^{\mathrm{eff},z}(t)$ can be expressed as
\begin{equation}
\dot{h}_i^{\mathrm{eff},z}(t)=\sum_{\substack{j \\ (j\neq i)}}\alpha_{ij}(\{m_k^z\},t)\dot{m}_j^z(t)+\beta_i(\{m_k^z\},t), 
\end{equation}
where $\alpha_{ij}(\{m_k^z\},t)$ is a function of $\{m_k^z\}$ ($k\neq i,j$) and time $t$, and $\beta_i(\{m_k^z\},t)$ is a function of $\{m_k^z\}$ ($k\neq i$) and time $t$. 
Then, the simultaneous equations of motion (\ref{Eq.EOM.detail}) become
\begin{equation}
\begin{pmatrix}
\bm{1} & \bm{0} & \bm{A}^{xz} \\
\bm{0} & \bm{1} & \bm{0} \\
\bm{0} & \bm{0} & \bm{1}-\bm{A}^{zz}
\end{pmatrix}
\begin{pmatrix}
\dot{\bm{m}}^x \\
\dot{\bm{m}}^y \\
\dot{\bm{m}}^z
\end{pmatrix}
=
\begin{pmatrix}
\bm{B}^x \\
\bm{B}^y \\
\bm{B}^z
\end{pmatrix},
\end{equation}
where ${^t(\dot{\bm{m}}^x\dot{\bm{m}}^y\dot{\bm{m}}^z)}$ is the vector ${^t(\dot{m}_1^x\cdots\dot{m}_N^x\dot{m}_1^y\cdots\dot{m}_N^y\dot{m}_1^z\cdots\dot{m}_N^z)}$, $\bm{1}$ is the $N\times N$ identity submatrix, $\bm{0}$ is the $N\times N$ zero submatrix, $\bm{A}^{xz}$ and $\bm{A}^{zz}$ are the $N\times N$ submatrices with the matrix elements
\begin{equation}
A_{ij}^{xz}=(1-\delta_{ij})\frac{m_i^z(t)h_i^{\mathrm{eff},x}(t)\alpha_{ij}(\{m_k^z\},t)}{|\bm{h}_i^\mathrm{eff}(t)|^2},
\end{equation}
and
\begin{equation}
A_{ij}^{zz}=(1-\delta_{ij})\frac{m_i^x(t)h_i^{\mathrm{eff},x}(t)\alpha_{ij}(\{m_k^z\},t)}{|\bm{h}_i^\mathrm{eff}(t)|^2},
\end{equation}
and $^t(\bm{B}^x\bm{B}^y\bm{B}^z)={^t}(B_1^x\cdots B_N^xB_1^y\cdots B_N^yB_1^z\cdots B_N^z)$ is the vector with the elements
\begin{equation}
\left\{
\begin{aligned}
&B_i^x=2m_i^y(t)h_i^{\mathrm{eff},z}(t)+\frac{m_i^z(t)h_i^{\mathrm{eff},z}(t)\dot{h}_i^{\mathrm{eff},x}(t)-m_i^z(t)h_i^{\mathrm{eff},x}(t)\beta_i(\{m_k^z\},t)}{|\bm{h}_i^\mathrm{eff}(t)|^2}, \\
&B_i^y=2m_i^z(t)h_i^{\mathrm{eff},x}(t)-2m_i^x(t)h_i^{\mathrm{eff},z}(t), \\
&B_i^z=-\frac{m_i^x(t)h_i^{\mathrm{eff},z}(t)\dot{h}_i^{\mathrm{eff},x}(t)-m_i^x(t)h_i^{\mathrm{eff},x}(t)\beta_i(\{m_k^z\},t)}{|\bm{h}_i^\mathrm{eff}(t)|^2}-2m_i^y(t)h_i^{\mathrm{eff},x}(t).
\end{aligned}
\right.
\end{equation}
Now we can express $\{\dot{\bm{m}}_i\}$ as a function of $\{\bm{m}_i\}$, that is, 
\begin{equation}
\begin{pmatrix}
\dot{\bm{m}}^x \\
\dot{\bm{m}}^y \\
\dot{\bm{m}}^z
\end{pmatrix}
=
\begin{pmatrix}
\bm{1} & \bm{0} & -\bm{A}^{xz}(\bm{1}-\bm{A}^{zz})^{-1} \\
\bm{0} & \bm{1} & \bm{0} \\
\bm{0} & \bm{0} & (\bm{1}-\bm{A}^{zz})^{-1}
\end{pmatrix}
\begin{pmatrix}
\bm{B}^x \\
\bm{B}^y \\
\bm{B}^z
\end{pmatrix}.
\end{equation}
This equation is just a $3N$-dimensional first-order differential equation, and thus it is easy to solve numerically. 
Note that the main computational cost of this algorithm arises from the calculation of $(\bm{1}-\bm{A}^{zz})^{-1}$, and thus the complexity of this algorithm is given by $\mathcal{O}(N^3)$. 

%
%
\section{Iterative approach}
%
%
\subsection{\label{Sec.IQAA}Iterative quantum adiabatic algorithm}

We explain an iterative quantum adiabatic algorithm to schedule inhomogeneous driving which enables us to avoid a small energy gap during adiabatic transformation~\cite{Dickson2012}. 
Suppose that the $n$th eigensector of the problem Hamiltonian corresponds to local or grobal minimum of energy and the Hamming distance among the $n$th eigenstates is more than $2$, that is, the equality $\hat{P}_n\hat{\mathcal{H}}_V\hat{P}_n=0$ holds, where $\hat{P}_n$ is the projection operator of the $n$th eigensector. 
Then, the first order perturbative correction to energy under a small transverse field is zero. 
The second order perturbative corrections to energy under a small transverse field are given by the eigenvalues of
\begin{equation}
\lambda^2\sum_{m(\neq n)}\frac{\hat{P}_n\hat{\mathcal{H}}_V\hat{P}_m\hat{\mathcal{H}}_V\hat{P}_n}{E_n^{(0)}-E_m^{(0)}},
\label{Eq.2nd.corr}
\end{equation}
where $E_n^{(0)}$ is the $n$th unperturbed eigenvalue of the problem Hamiltonian and $\lambda=[1-g(t/T)]$ with $t/T\approx1$. 
The second order perturbative corrections are negative because the numerator $\hat{P}_n\hat{\mathcal{H}}_V\hat{P}_m\hat{\mathcal{H}}_V\hat{P}_n$ is always positive and the denominator $E_n^{(0)}-E_m^{(0)}$, where $E_n^{(0)}$ is the energy of local or global minimum and $E_m^{(0)}$ is that of a state obtained by single bit flip from local or global minimum, is always negative. 
Suppose that an eigenvector of Eq.~(\ref{Eq.2nd.corr}) is given by
\begin{equation}
|n,\mu\rangle=\sum_\nu c_{n,\mu}^\nu|n,\nu\rangle_0,
\end{equation}
where the coefficient $c_{n,\mu}^\nu$ satisfies $\sum_\nu|c_{n,\mu}^\nu|^2=1$ and $|n,\nu\rangle_0$ is the unperturbed eigenstate of the problem Hamiltonian. 
Then, the second order perturbative correction is given by
\begin{equation}
E_{n,\mu}^{(2)}=\frac{\lambda^2}{4}\sum_{\{i,j,\kappa,\nu\}}\frac{\Delta_i\Delta_jc_{n,\mu}^{\kappa\ast}c_{n,\mu}^\nu}{\Delta E_{n,i}},
\end{equation}
where the summation $\sum_{\{i,j,\kappa,\nu\}}$ is taken over so that the state $|n,\kappa\rangle_0$ is obtained by flipping the $i$th and the $j$th bits of the state $|n,\nu\rangle_0$. 
Here, $\Delta E_{n,i}\equiv E_n^{(0)}-E_m^{(0)}$, where $E_m^{(0)}$ is the unperturbed eigenenergy of the state obtained by flipping the $i$th (or the $j$th) bit of the state $|n,\nu\rangle_0$. 
If these corrections to local minima are large and that to the global minimum is small, small energy gaps between the global minimum and these local minima may appear. 
Therefore, we want to reduce these corrections to local minima except for that to the global minimum.

In Ref.~\cite{Dickson2012}, this correction term is written as $E_{n,\mu}^{(2)}=\sum_i\Delta_i\mu_i$ and $\mu_i$ is calculated by sampling under some assumptions and some approximations (see Ref.~\cite{Dickson2012} for details). 
To minimize this correction term with keeping $(\prod_i\Delta_i)^{1/N}$ being constant suggests to set $\Delta_i\propto\mu_i^{-1}$. 
In Ref.~\cite{Dickson2012}, a transverse field was updated as $\Delta_{i,\mathrm{old}}\to\Delta_{i,\mathrm{new}}\propto\Delta_{i,\mathrm{old}}^{1-\beta}\mu_i^{-\beta}$, where $\beta=1/(\kappa+1)$ and $\kappa$ is the number of iteration.  
Under this iteration rule, mitigation of small energy gaps in the maximum independent set problem was found~\cite{Dickson2012}. 
Later we will show that a similar way of iteration can also remove energy crossings in the classical model of the quantum adiabatic algorithm.

%
%
\subsection{\label{Sec.ite.classical}Iterative classical superadiabatic algorithm}
In the classical superadiabatic algorithm, the annealing process always results in one of the local minima of the problem Hamiltonian except for cases showing critical divergence. 
We therefore calculate the expectation value of the matrix giving the second order perturbative corrections to energy (\ref{Eq.2nd.corr}) by using this metastable state. 
It is given by
\begin{equation}
E_\mathrm{MS}^{(2)}=\frac{\lambda^2}{4}\sum_i\frac{\Delta_i^2}{\Delta E_{\mathrm{MS},i}},
\label{Eq.classical.ene.corr}
\end{equation}
where $\Delta E_{\mathrm{MS},i}$ is the energy difference between the energy of the metastable state and that of a state obtained by flipping the $i$th bit of the metastable state. 
Note that this quantity is also negative. 
In order to avoid formation of energy crossings, we want to reduce this quantity. 
By using the Lagrange multiplier for the condition, $(\prod_i\Delta_i)^{1/N}=\mathrm{const.}$, minimization of the quantity (\ref{Eq.classical.ene.corr}) suggests the following inhomogeneous transverse field
\begin{equation}
\Delta_i^2=|\Delta E_{\mathrm{MS},i}|\left(\prod_{i=1}^N\frac{\Delta_i^2}{|\Delta E_{\mathrm{MS},i}|}\right)^{1/N},
\end{equation}
that is, $\Delta_i^2\propto|\Delta E_{\mathrm{MS},i}|$. 
Note that if we update a transverse field as $\Delta_i=\sqrt{|\Delta E_{\mathrm{MS},i}|}$, the strength of updated transverse fields may always be the same or may form a loop among some certain values. 
In order to avoid these situations, we update a transverse field as
\begin{equation}
\Delta_{i,\mathrm{old}}\to\Delta_{i,\mathrm{new}}=\frac{|\Delta E_{\mathrm{MS},i}|}{\Delta_{i,\mathrm{old}}}. 
\label{Eq.ite.method}
\end{equation}
Here, the strength of the updated transverse field depends on that of previous transverse fields, and thus we can avoid trivial loops of iteration. 
Moreover, if this update converges at a certain value, i.e., the equality $\Delta_{i,\mathrm{new}}=\Delta_{i,\mathrm{old}}$ holds after some iteration, it satisfies $\Delta_{i,\mathrm{new}}=\sqrt{|\Delta E_{\mathrm{MS},i}|}$. 
Finally we mention physical meaning of this update rule. 
The energy increase $|\Delta E_{\mathrm{MS},i}|$ is the energy difference between the energy of the metastable state and that of a state obtained by flipping the $i$th bit of the metastable state, and thus it becomes large if the $i$th bit involves many other bits. 
Therefore, the update rule (\ref{Eq.ite.method}) just implies to apply a strong local transverse field to a ``hub" bit. 
We will see that this simple update rule can resolve a large amount of the hard instances discussed in Sec.~\ref{Sec.QAA}.

%
%
\section{\label{Sec.results}Numerical simulation}

%
%
\subsection{\label{Sec.planted}Two planted solutions}
\begin{figure}
\includegraphics[width=10cm]{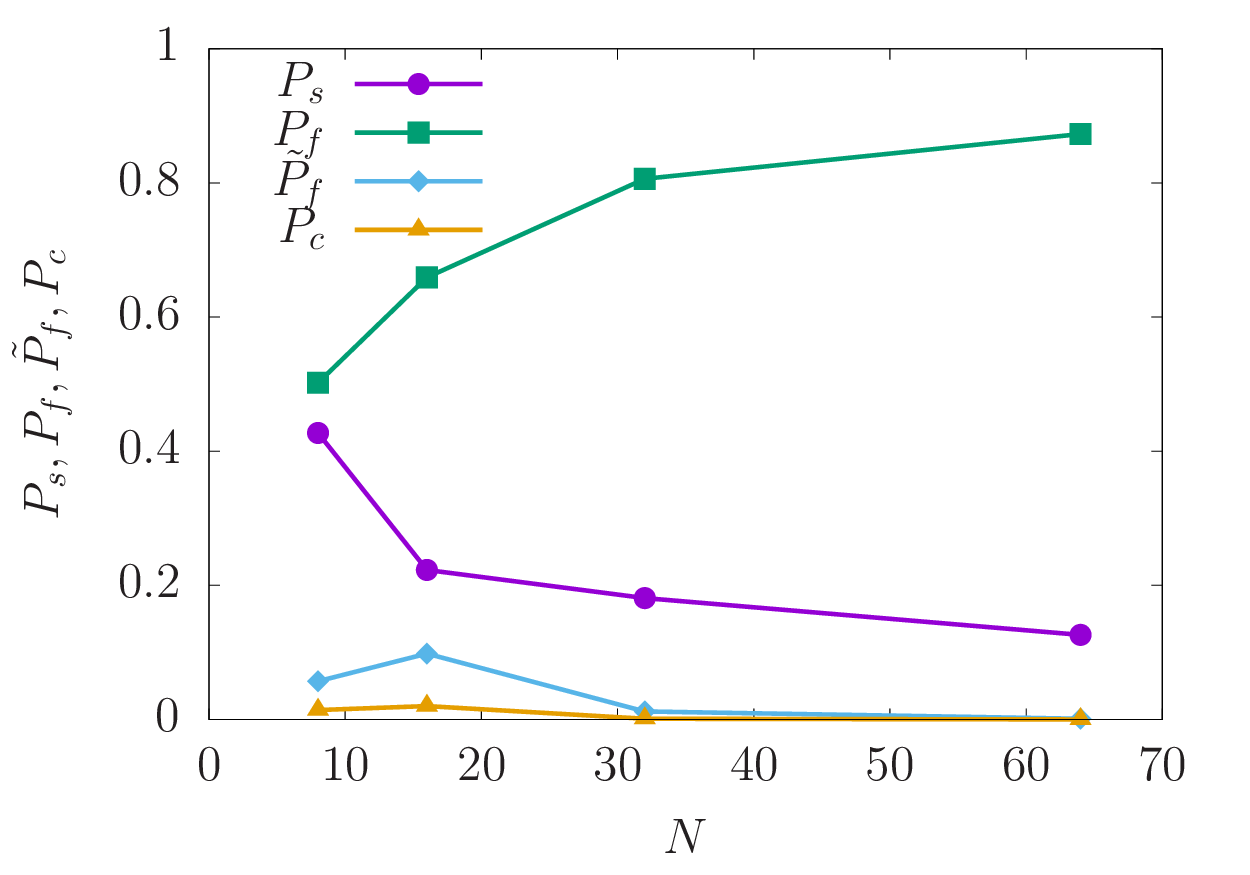}
\caption{\label{Fig.SAT2gs_plantedsol}Rate of the cases where the process results in (i, purple circles) success in finding the planted solution corresponding to the ground state, (ii, green squares) failure in finding the planted solution corresponding to the ground state but resulting in the other planted solution, (iii, cyan diamonds) failure in finding the planted solutions and resulting in a non-planted solution corresponding to a metastable state, and (iv, orange triangles) failure in finding any approximate solution because of criticality. The horizontal axis is the system size $N$ and the vertical axis is those rate. }
\end{figure}

First, we check that instances with two planted solutions, $00\cdots0$ and $11\cdots1$, discussed in Ref.~\cite{Farhi2011} (see also Sec.~\ref{Sec.QAA}), which show small energy gaps during adiabatic transformation in the quantum adiabatic algorithm, also show energy crossings in the classical model of the quantum adiabatic algorithm, i.e., these instances are actually hard even in the classical regime. 
In order to show it, we apply the classical superadiabatic algorithm discussed in Sec.~\ref{Sec.CSA} to those instances. 
Here, for each number of bits $N$, we generate $1000$ instances and we use a uniform transverse field $\Delta_i=4$. 
Each outcome is classified into four cases, i.e., (i) success in finding the planted solution corresponding to the ground state, (ii) failure in finding the planted solution corresponding to the ground state but resulting in the other planted solution, (iii) failure in finding the planted solutions and resulting in a non-planted solution corresponding to a metastable state, and (iv) failure in finding any approximate solution because of criticality. 
Here, energy crossings appear in the cases (ii) and (iii), and these may happen in the case (iv). 
We study the rate of these cases among 1000 instances and plot them with respect to the system size in Fig.~\ref{Fig.SAT2gs_plantedsol}, where $P_s$, $P_f$, $\tilde{P}_f$, and $P_c$ are the rate of the cases (i), (ii), (iii), and (iv), respectively. 
We find that the rate of failure $P_f$ (the rate of success $P_s$), i.e., the case (ii) [(i)], increases (decreases) along with the system size $N$, and thus we successfully generate hard instances for large $N$ even in the classical regime although this generation method is based on quantum mechanics. 
Note that there are some instances resulting in non-planted solutions or in critical divergence for small system size $N$, but these disappear for large $N$, and thus we conclude that these instances appear bacause of unwanted satisfying assignments (see Sec.~\ref{Sec.QAA} and Ref.~\cite{Farhi2011}).

Our method can also be used to study other properties of instances. 
Here we study when those energy crossings happen during the annealing processes of those hard instances. 
In order to study it, we first prepare two planted solutions, $00\cdots0$ and $11\cdots1$, and then we simulate rewinding processes from those states with our method. 
By calculating energy during those processes, we can directly observe energy crossings. 
Here, in order to compare with the previous work studying in the quantum regime~\cite{Farhi2011}, we set a uniform transverse field $\Delta_i=1$. 
On average, energy crossings happen when $g(t/T)=0.21\pm0.12$ for $N=8$, $g(t/T)=0.30\pm0.12$ for $N=16$, $g(t/T)=0.29\pm0.11$ for $N=32$, and $g(t/T)=0.29\pm0.10$ for $N=64$. 
These results differ from the observation and the perturbative expansion analysis in the quantum case, where $g(t/T)\approx0.42$ for $N=16$ and $g(t/T)\approx0.49$ for $N=150$~\cite{Farhi2011}, whereas some instances show such values even in the classical case. 
However, this difference is not very surprising because quantumness of a state would be almost maximum around the middle of the annealing process and properties of stationary states in the classical model should differ from properties of eigenstates in the quantum adiabatic algorithm. 
The important point is that hard instances showing small energy gaps in the quantum adiabatic algorithm are still hard for the classical model of the quantum adiabatic algorithm, that is, these instances show energy crossings in the classical model of the quantum adiabatic algorithm.

%
%
\subsection{\label{Sec.resolve.hard}Resolving hard instances}
\begin{figure}
\includegraphics[width=10cm]{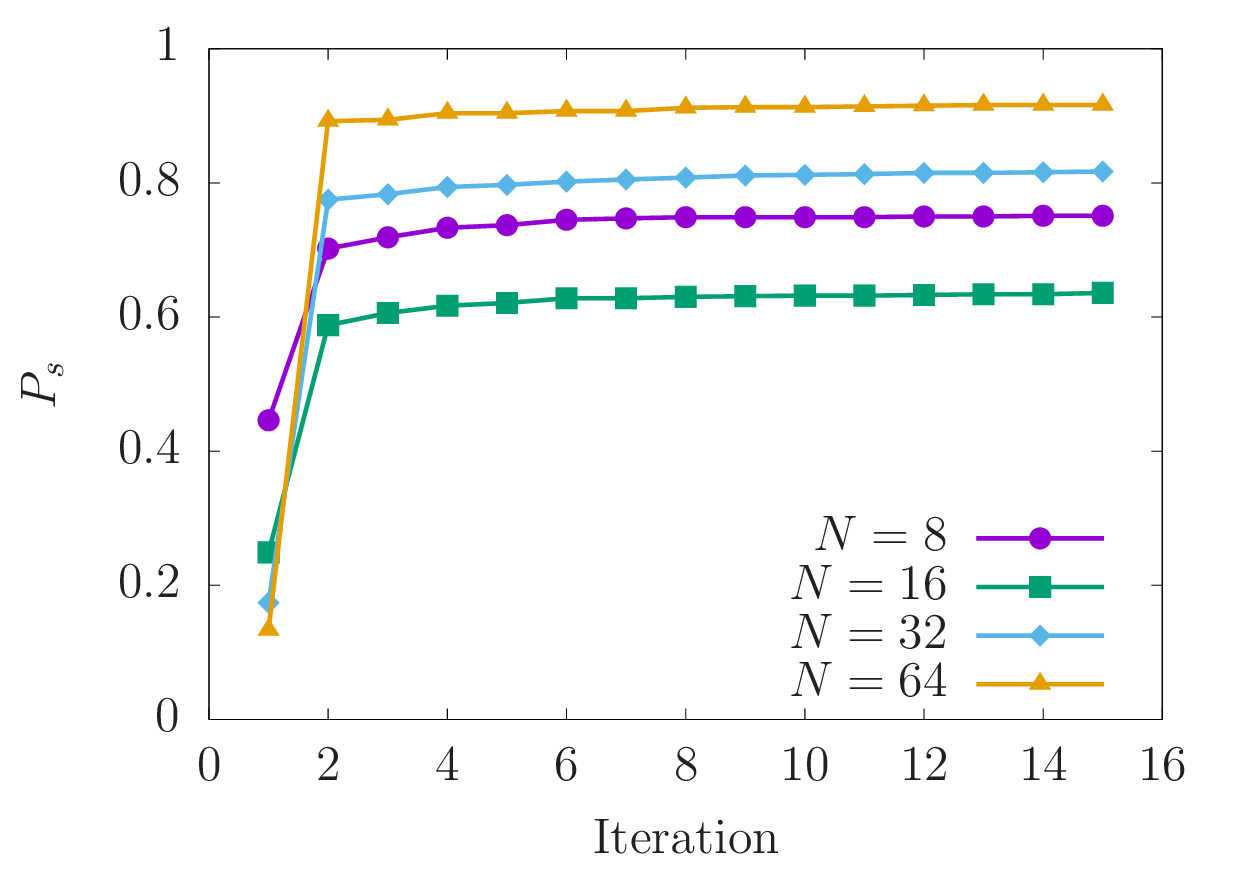}
\caption{\label{Fig.itepersuc}Rate of success with respect to the number of iteration. The horizontal axis is the number of iteration and the vertical axis is the rate of success $P_s$. The system size is $N=8$ (purple circles), $16$ (green squares), $32$ (cyan diamonds), and $64$ (orange triangles). }
\end{figure}

Now we try to resolve these hard instances by using the iterative approach discussed in Sec.~\ref{Sec.ite.classical}. 
Here, for a given instance, we update transverse fields 15 times at most according to Eq.~(\ref{Eq.ite.method}) and rescale them so that $\Delta_i\in[2,10]$. 
Note that we use a uniform transverse field $\Delta_i=4$ as the initial setup. 
We plot how the rate of success among 1000 instances changes along with iteration in Fig.~\ref{Fig.itepersuc}. 
We find that the first iteration dramatically improves the rate of success and the following iteration does not make big difference. 
Moreover, the number of instances that are resolved after iteration increases when the system size becomes larger.

\begin{figure}
\includegraphics[width=10cm]{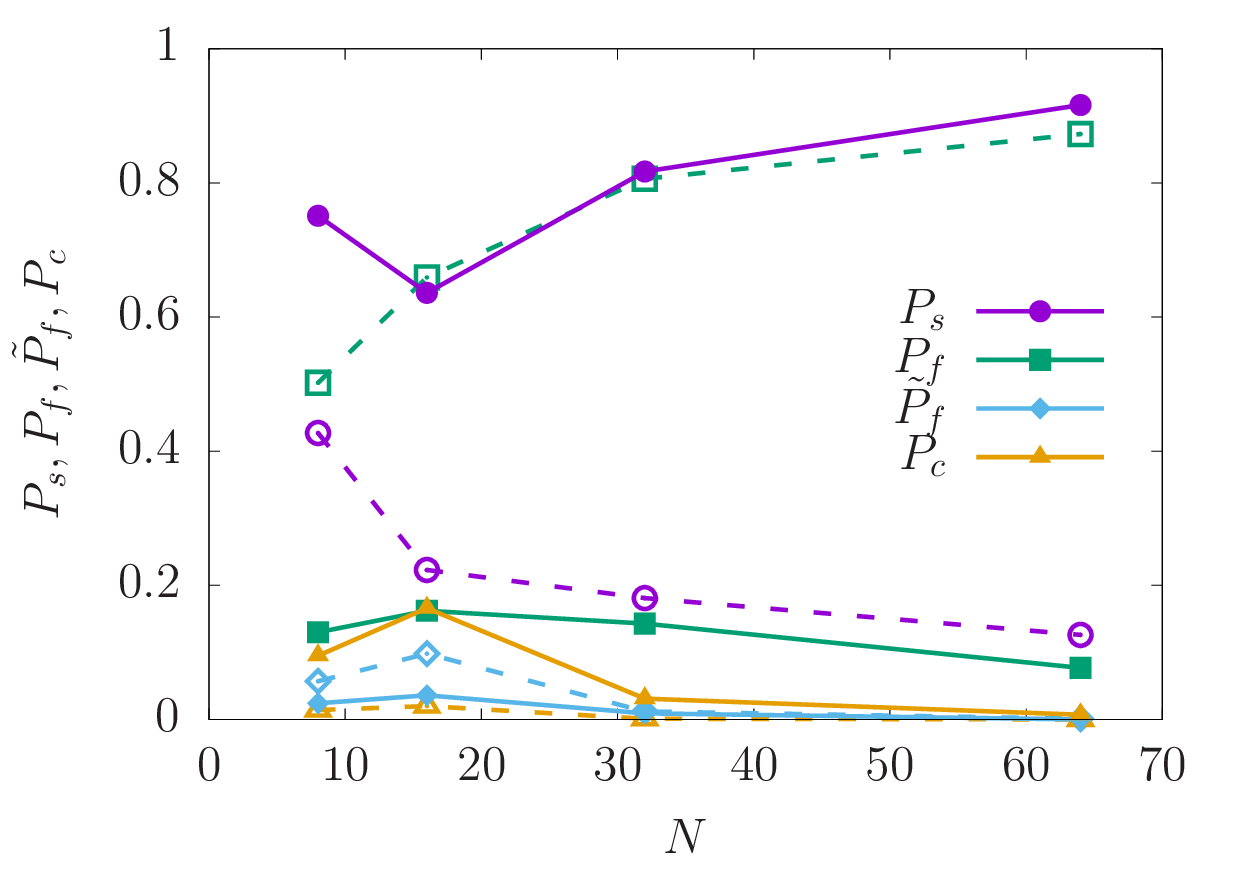}
\caption{\label{Fig.itesffcf2t10}Rate of the cases where the process results in (i, purple circles) success in finding the planted solution corresponding to the ground state, (ii, green squares) failure in finding the planted solution corresponding to the ground state but resulting in the other planted solution, (iii, cyan diamonds) failure in finding the planted solutions and resulting in a non-planted solution corresponding to a metastable state, and (iv, orange triangles) failure in finding any approximate solution because of criticality. Those after 15 iteration are plotted by solid lines with closed symbols and those for a uniform transverse field are supplementally plotted by dashed lines with open symbols. The horizontal axis is the system size $N$ and the vertical axis is the rate of these cases. }
\end{figure}

\begin{figure}
\includegraphics[width=10cm]{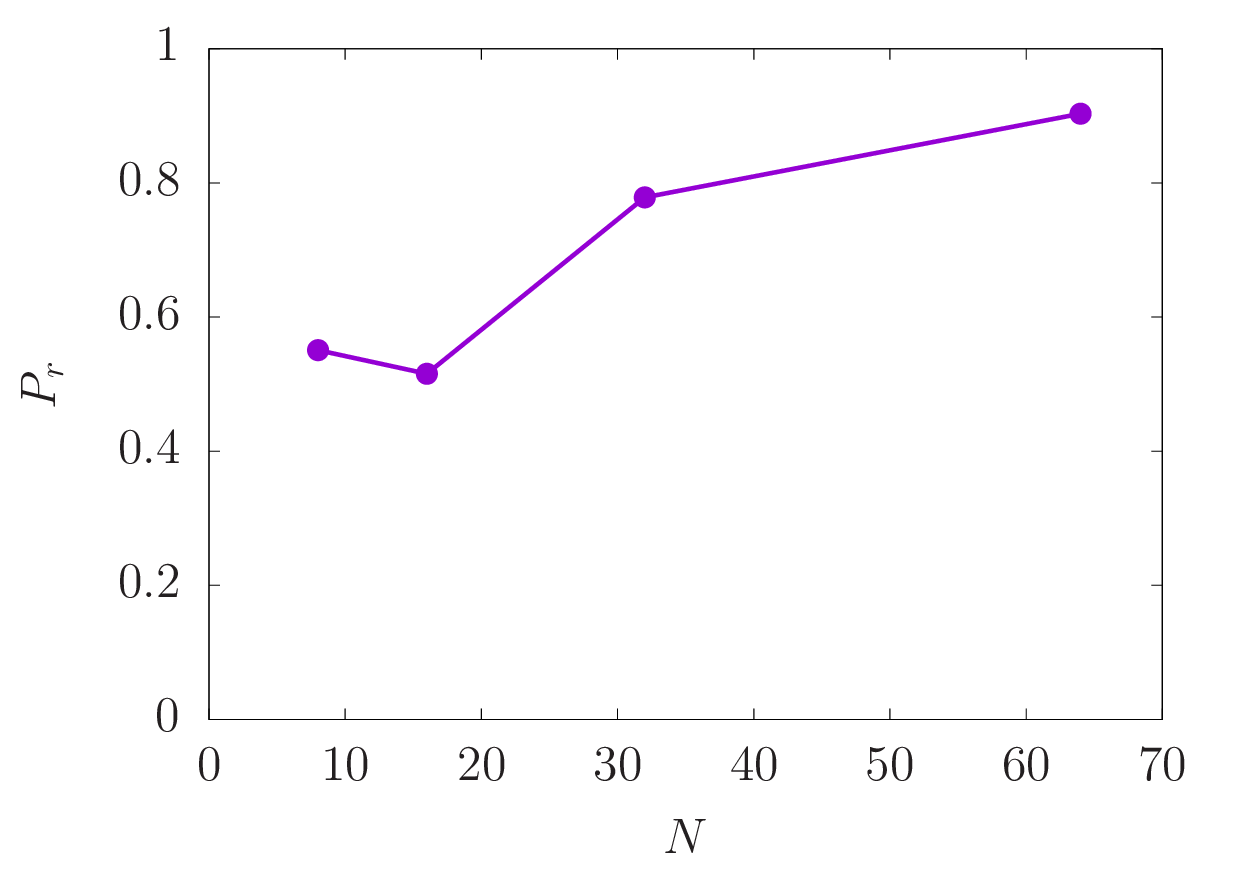}
\caption{\label{Fig.resolve_hard_rate}Rate of resolution by iteration. The horizontal axis is the system size $N$ and the vertical axis is the rate of resolution $P_r$. }
\end{figure}

We also check the other rate of the cases (i) - (iv) after 15 iteration and plot those in Fig.~\ref{Fig.itesffcf2t10}. 
Together with Fig.~\ref{Fig.itepersuc}, we find that the rate of success (failure) drastically increases (decreases). 
In particular, for lager system size, the present approach works better (except for $N=16$). 
Indeed, for $N=64$, the rate of success reaches more than 90\%, whereas it is less than 20\% under a uniform transverse field. 
We also calculate the rate of resolution by iteration. 
It is given by
\begin{equation}
P_r=\frac{P_{s,\mathrm{tot}}-P_{s,\mathrm{uni}}}{1-P_{s,\mathrm{uni}}}=\frac{P_{s,\mathrm{ite}}}{1-P_{s,\mathrm{uni}}},
\end{equation}
where $P_{s,\mathrm{tot}}$ is the rate of success after 15 iteration (purple closed circles in Fig.~\ref{Fig.itesffcf2t10}), $P_{s,\mathrm{uni}}$ is the rate of success for a uniform transverse field (purple open circles in Fig.~\ref{Fig.itesffcf2t10}), and $P_{s,\mathrm{ite}}$ is the rate of success by iteration, $P_{s,\mathrm{ite}}=P_{s,\mathrm{tot}}-P_{s,\mathrm{uni}}$, and plot it in Fig.~\ref{Fig.resolve_hard_rate}. 
We find that the rate of resolution by iteration also reaches more than 90\% for the system size $N=64$. 
Note that the rest 10\% of instances may truly hard instances or may be resolved by other ways of iteration. 
At present, we have neither answer nor (efficient) resolution, but this point should be studied in the future.

%
%
\section{\label{Sec.discuss}Discussion}
Inspired by the quantum adiabatic algorithm~\cite{Farhi2000,Farhi2001,Kadowaki1998}, which is a quantum algorithm, and by the iterative quantum adiabatic algorithm~\cite{Dickson2012}, which is a hybrid quantum-classical algorithm, we considered the classical and superadiabatic version of the iterative quantum adiabatic algorithm and application of it to a certain class of hard instances of the 3-SAT problem. 
Our algorithm is completely classical and does not use any stochastic approach to simulate quantum dynamics, and thus for a given problem and under given initial conditions we can obtain a result in a deterministic way. 
This deterministic property enables us to study problem instances themselves without stochastic uncertainty. 
In addition, we used the exact theory of shortcuts to adiabaticity for classical spin systems~\cite{Hatomura2018b}, and thus our algorithm is independent of the annealing time and it has ability to solve easy instances, where no energy crossing appears, with the probability 100\%. 
Therefore, we can also discuss instances themselves without shortage of adiabaticity. 
Note that for quantum spin systems only a few exact construction of shortcuts is known~\cite{DelCampo2012}, and thus for the quantum adiabatic algorithm we have to rely on approximate construction of shortcuts which induce some deviation from adiabatic transformation~\cite{Takahashi2017,Hatomura2017,Ozguler2018,Hartmann2019a}. 
This is one of the reasons why we considered the classical model. 
By using our algorithm, we checked that a certain class of hard instances of the 3-SAT problem in the quantum adiabatic algorithm, where small energy gaps appear, is also hard in the classical model, that is, energy crossings happen. 
Moreover, for such hard instances, we found that more than 90\% of these 64-bits hard instances can be resolved within a few iteration of inhomogeneous driving, whereas resolution of hard instances is exponentially hard if we use random inhomogenous driving~\cite{Hatomura2018b}.

Here we revisit the concept of adiabaticity and that of energy (anti)crossings in quantum and classical mechanics. 
In quantum mechanics, adiabatic time evolution is dynamics that tracks instantaneous eigenstates of a slowly varying Hamiltonian. 
Eigenvalues of those instantaneous eigenstates sometimes cross each other. 
However, because of quantumness, they usually form an avoided crossing, i.e., an energy gap. 
Therefore, starting from the ground state, a given system can remain in the ground state if change of its Hamiltonian is slow enough. 
This property ensures success in solving a given problem by using the quantum adiabatic algorithm although it may require exponentially long time. 
In classical mechanics, adiabatic time evolution is dynamics that conserves volume of an equal energy surface in phase space or that tracks an instantaneous stationary state of a slowly varying Hamiltonian, where an equal energy surface is a point in phase space. 
Energy of instantaneous stationary states may also cross each other. 
In contrast to quantum mechanics, they do not form an avoided crossing, and thus a system cannot remain in the ground state, but it becomes a metastable state at a crossing point even if change of its Hamiltonian is infinitely slow. 
This is the main drawback of the classical model of the quantum adiabatic algorithm. 
However, such instances showing energy crossings in the classical model of the quantum adiabatic algorithm may also cause exponentially small energy gaps in the quantum adiabatic algorithm. 
As an example, instances discussed in the present paper showing energy crossings in the classical model of the quantum adiabatic algorithm actually cause small energy gaps in the quantum adiabatic algorithm~\cite{Farhi2011}. 
In this case, for large systems, hardness of the quantum adiabatic algorithm would be similar to that of the classical model of quantum adiabatic algorithm, i.e., quantum and classical dynamics would show similar behavior around energy (anti)crossings for realistic finite time processes.

Of course, there is difference in possible dynamics of the quantum adiabatic algorithm and that of the classical model of the quantum adiabatic algorithm. 
A classical spin system consisting of three-dimensional unit vectors can be described by a product state of two-level systems, that is, dynamics of a classical spin system is limited to subspace, where the corresponding quantum spin system is given by a product state. 
Therefore, if dynamics of the quantum adiabatic algorithm involves large amount of entanglement, dynamics of the classical model of the quantum adiabatic algorithm is far from it. 
Indeed, we found that place of energy crossings in the classical regime differs from that of small energy gaps in the quantum regime. 
Therefore, dynamics should differ between the classical and quantum cases around the middle of annealing processes. 
Nevertheless, energy crossings can be removed by using a similar way of iteration for inhomogeneous driving, which mitigates small energy gaps in the quantum adiabatic algorithm, even in the classical model of the quantum adiabatic algorithm.

In conclusion, our method elucidates both similarity and difference between the quantum adiabatic algorithm and its classical model apart from stochastic uncertainty and shortage of adiabaticity. 
Even if there is difference, methods derived within quantum perturbation theory can work in the classical model. 
Similarly, methods derived within classical theory or properties found in the classical model could also work in the quantum adiabatic algorithm. 
We believe that, in addition to the iterative approach for inhomogeneous driving, we can obtain many insight to improve performance of the quantum adiabatic algorithm from the classical model.

\bibliography{inhomoAQCbib}

\begin{thebibliography}{27}%
\makeatletter
\providecommand \@ifxundefined [1]{%
 \@ifx{#1\undefined}
}%
\providecommand \@ifnum [1]{%
 \ifnum #1\expandafter \@firstoftwo
 \else \expandafter \@secondoftwo
 \fi
}%
\providecommand \@ifx [1]{%
 \ifx #1\expandafter \@firstoftwo
 \else \expandafter \@secondoftwo
 \fi
}%
\providecommand \natexlab [1]{#1}%
\providecommand \enquote  [1]{``#1''}%
\providecommand \bibnamefont  [1]{#1}%
\providecommand \bibfnamefont [1]{#1}%
\providecommand \citenamefont [1]{#1}%
\providecommand \href@noop [0]{\@secondoftwo}%
\providecommand \href [0]{\begingroup \@sanitize@url \@href}%
\providecommand \@href[1]{\@@startlink{#1}\@@href}%
\providecommand \@@href[1]{\endgroup#1\@@endlink}%
\providecommand \@sanitize@url [0]{\catcode `\\12\catcode `\$12\catcode
  `\&12\catcode `\#12\catcode `\^12\catcode `\_12\catcode `\%12\relax}%
\providecommand \@@startlink[1]{}%
\providecommand \@@endlink[0]{}%
\providecommand \url  [0]{\begingroup\@sanitize@url \@url }%
\providecommand \@url [1]{\endgroup\@href {#1}{\urlprefix }}%
\providecommand \urlprefix  [0]{URL }%
\providecommand \Eprint [0]{\href }%
\providecommand \doibase [0]{http://dx.doi.org/}%
\providecommand \selectlanguage [0]{\@gobble}%
\providecommand \bibinfo  [0]{\@secondoftwo}%
\providecommand \bibfield  [0]{\@secondoftwo}%
\providecommand \translation [1]{[#1]}%
\providecommand \BibitemOpen [0]{}%
\providecommand \bibitemStop [0]{}%
\providecommand \bibitemNoStop [0]{.\EOS\space}%
\providecommand \EOS [0]{\spacefactor3000\relax}%
\providecommand \BibitemShut  [1]{\csname bibitem#1\endcsname}%
\let\auto@bib@innerbib\@empty
\bibitem [{\citenamefont {Farhi}\ \emph {et~al.}(2000)\citenamefont {Farhi},
  \citenamefont {Goldstone}, \citenamefont {Gutmann},\ and\ \citenamefont
  {Sipser}}]{Farhi2000}%
  \BibitemOpen
  \bibfield  {author} {\bibinfo {author} {\bibfnamefont {E.}~\bibnamefont
  {Farhi}}, \bibinfo {author} {\bibfnamefont {J.}~\bibnamefont {Goldstone}},
  \bibinfo {author} {\bibfnamefont {S.}~\bibnamefont {Gutmann}}, \ and\
  \bibinfo {author} {\bibfnamefont {M.}~\bibnamefont {Sipser}},\ }\href
  {http://arxiv.org/abs/quant-ph/0001106} {\  (\bibinfo {year} {2000})},\
  \Eprint {http://arxiv.org/abs/0001106} {arXiv:0001106 [quant-ph]}
  \BibitemShut {NoStop}%
\bibitem [{\citenamefont {Farhi}\ \emph {et~al.}(2001)\citenamefont {Farhi},
  \citenamefont {Goldstone}, \citenamefont {Gutmann}, \citenamefont {Lapan},
  \citenamefont {Lundgren},\ and\ \citenamefont {Preda}}]{Farhi2001}%
  \BibitemOpen
  \bibfield  {author} {\bibinfo {author} {\bibfnamefont {E.}~\bibnamefont
  {Farhi}}, \bibinfo {author} {\bibfnamefont {J.}~\bibnamefont {Goldstone}},
  \bibinfo {author} {\bibfnamefont {S.}~\bibnamefont {Gutmann}}, \bibinfo
  {author} {\bibfnamefont {J.}~\bibnamefont {Lapan}}, \bibinfo {author}
  {\bibfnamefont {A.}~\bibnamefont {Lundgren}}, \ and\ \bibinfo {author}
  {\bibfnamefont {D.}~\bibnamefont {Preda}},\ }\href {\doibase
  10.1126/science.1057726} {\bibfield  {journal} {\bibinfo  {journal}
  {Science}\ }\textbf {\bibinfo {volume} {292}},\ \bibinfo {pages} {472 }
  (\bibinfo {year} {2001})}\BibitemShut {NoStop}%
\bibitem [{\citenamefont {Kadowaki}\ and\ \citenamefont
  {Nishimori}(1998)}]{Kadowaki1998}%
  \BibitemOpen
  \bibfield  {author} {\bibinfo {author} {\bibfnamefont {T.}~\bibnamefont
  {Kadowaki}}\ and\ \bibinfo {author} {\bibfnamefont {H.}~\bibnamefont
  {Nishimori}},\ }\href {\doibase 10.1103/PhysRevE.58.5355} {\bibfield
  {journal} {\bibinfo  {journal} {Physical Review E}\ }\textbf {\bibinfo
  {volume} {58}},\ \bibinfo {pages} {5355} (\bibinfo {year}
  {1998})}\BibitemShut {NoStop}%
\bibitem [{\citenamefont {Lucas}(2014)}]{Lucas2014}%
  \BibitemOpen
  \bibfield  {author} {\bibinfo {author} {\bibfnamefont {A.}~\bibnamefont
  {Lucas}},\ }\href {\doibase 10.3389/fphy.2014.00005} {\bibfield  {journal}
  {\bibinfo  {journal} {Frontiers in Physics}\ }\textbf {\bibinfo {volume}
  {2}},\ \bibinfo {pages} {5} (\bibinfo {year} {2014})}\BibitemShut {NoStop}%
\bibitem [{\citenamefont {Kato}(1950)}]{Kato1950}%
  \BibitemOpen
  \bibfield  {author} {\bibinfo {author} {\bibfnamefont {T.}~\bibnamefont
  {Kato}},\ }\href {\doibase 10.1143/JPSJ.5.435} {\bibfield  {journal}
  {\bibinfo  {journal} {Journal of the Physical Society of Japan}\ }\textbf
  {\bibinfo {volume} {5}},\ \bibinfo {pages} {435} (\bibinfo {year}
  {1950})}\BibitemShut {NoStop}%
\bibitem [{\citenamefont {Farhi}\ \emph {et~al.}(2002)\citenamefont {Farhi},
  \citenamefont {Goldstone},\ and\ \citenamefont {Gutmann}}]{Farhi2002}%
  \BibitemOpen
  \bibfield  {author} {\bibinfo {author} {\bibfnamefont {E.}~\bibnamefont
  {Farhi}}, \bibinfo {author} {\bibfnamefont {J.}~\bibnamefont {Goldstone}}, \
  and\ \bibinfo {author} {\bibfnamefont {S.}~\bibnamefont {Gutmann}},\ }\href
  {http://arxiv.org/abs/quant-ph/0208135} {\  (\bibinfo {year} {2002})},\
  \Eprint {http://arxiv.org/abs/0208135} {arXiv:0208135 [quant-ph]}
  \BibitemShut {NoStop}%
\bibitem [{\citenamefont {Farhi}\ \emph {et~al.}(2011)\citenamefont {Farhi},
  \citenamefont {Goldstone}, \citenamefont {Gosset}, \citenamefont {Gutmann},
  \citenamefont {Meyer},\ and\ \citenamefont {Shor}}]{Farhi2011}%
  \BibitemOpen
  \bibfield  {author} {\bibinfo {author} {\bibfnamefont {E.}~\bibnamefont
  {Farhi}}, \bibinfo {author} {\bibfnamefont {J.}~\bibnamefont {Goldstone}},
  \bibinfo {author} {\bibfnamefont {D.}~\bibnamefont {Gosset}}, \bibinfo
  {author} {\bibfnamefont {S.}~\bibnamefont {Gutmann}}, \bibinfo {author}
  {\bibfnamefont {H.~B.}\ \bibnamefont {Meyer}}, \ and\ \bibinfo {author}
  {\bibfnamefont {P.}~\bibnamefont {Shor}},\ }\href {\doibase
  10.26421/QIC11.3-4} {\bibfield  {journal} {\bibinfo  {journal} {Quantum
  Information \& Computation}\ }\textbf {\bibinfo {volume} {11}},\ \bibinfo
  {pages} {181} (\bibinfo {year} {2011})}\BibitemShut {NoStop}%
\bibitem [{\citenamefont {Dickson}\ and\ \citenamefont
  {Amin}(2011)}]{Dickson2011}%
  \BibitemOpen
  \bibfield  {author} {\bibinfo {author} {\bibfnamefont {N.~G.}\ \bibnamefont
  {Dickson}}\ and\ \bibinfo {author} {\bibfnamefont {M.~H.~S.}\ \bibnamefont
  {Amin}},\ }\href {\doibase 10.1103/PhysRevLett.106.050502} {\bibfield
  {journal} {\bibinfo  {journal} {Physical Review Letters}\ }\textbf {\bibinfo
  {volume} {106}},\ \bibinfo {pages} {050502} (\bibinfo {year}
  {2011})}\BibitemShut {NoStop}%
\bibitem [{\citenamefont {Dickson}\ and\ \citenamefont
  {Amin}(2012)}]{Dickson2012}%
  \BibitemOpen
  \bibfield  {author} {\bibinfo {author} {\bibfnamefont {N.~G.}\ \bibnamefont
  {Dickson}}\ and\ \bibinfo {author} {\bibfnamefont {M.~H.}\ \bibnamefont
  {Amin}},\ }\href {\doibase 10.1103/PhysRevA.85.032303} {\bibfield  {journal}
  {\bibinfo  {journal} {Physical Review A}\ }\textbf {\bibinfo {volume} {85}},\
  \bibinfo {pages} {032303} (\bibinfo {year} {2012})}\BibitemShut {NoStop}%
\bibitem [{\citenamefont {Lanting}\ \emph {et~al.}(2017)\citenamefont
  {Lanting}, \citenamefont {King}, \citenamefont {Evert},\ and\ \citenamefont
  {Hoskinson}}]{Lanting2017}%
  \BibitemOpen
  \bibfield  {author} {\bibinfo {author} {\bibfnamefont {T.}~\bibnamefont
  {Lanting}}, \bibinfo {author} {\bibfnamefont {A.~D.}\ \bibnamefont {King}},
  \bibinfo {author} {\bibfnamefont {B.}~\bibnamefont {Evert}}, \ and\ \bibinfo
  {author} {\bibfnamefont {E.}~\bibnamefont {Hoskinson}},\ }\href {\doibase
  10.1103/PhysRevA.96.042322} {\bibfield  {journal} {\bibinfo  {journal}
  {Physical Review A}\ }\textbf {\bibinfo {volume} {96}},\ \bibinfo {pages}
  {042322} (\bibinfo {year} {2017})}\BibitemShut {NoStop}%
\bibitem [{\citenamefont {Adame}\ and\ \citenamefont
  {McMahon}(2018)}]{Adame2018}%
  \BibitemOpen
  \bibfield  {author} {\bibinfo {author} {\bibfnamefont {J.~I.}\ \bibnamefont
  {Adame}}\ and\ \bibinfo {author} {\bibfnamefont {P.~L.}\ \bibnamefont
  {McMahon}},\ }\href {http://arxiv.org/abs/1806.11091} {\  (\bibinfo {year}
  {2018})},\ \Eprint {http://arxiv.org/abs/1806.11091} {arXiv:1806.11091}
  \BibitemShut {NoStop}%
\bibitem [{\citenamefont {Rams}\ \emph {et~al.}(2016)\citenamefont {Rams},
  \citenamefont {Mohseni},\ and\ \citenamefont {del Campo}}]{Rams2016a}%
  \BibitemOpen
  \bibfield  {author} {\bibinfo {author} {\bibfnamefont {M.~M.}\ \bibnamefont
  {Rams}}, \bibinfo {author} {\bibfnamefont {M.}~\bibnamefont {Mohseni}}, \
  and\ \bibinfo {author} {\bibfnamefont {A.}~\bibnamefont {del Campo}},\ }\href
  {\doibase 10.1088/1367-2630/aa5079} {\bibfield  {journal} {\bibinfo
  {journal} {New Journal of Physics}\ }\textbf {\bibinfo {volume} {18}},\
  \bibinfo {pages} {123034} (\bibinfo {year} {2016})}\BibitemShut {NoStop}%
\bibitem [{\citenamefont {Mohseni}\ \emph {et~al.}(2018)\citenamefont
  {Mohseni}, \citenamefont {Strumpfer},\ and\ \citenamefont
  {Rams}}]{Mohseni2018}%
  \BibitemOpen
  \bibfield  {author} {\bibinfo {author} {\bibfnamefont {M.}~\bibnamefont
  {Mohseni}}, \bibinfo {author} {\bibfnamefont {J.}~\bibnamefont {Strumpfer}},
  \ and\ \bibinfo {author} {\bibfnamefont {M.~M.}\ \bibnamefont {Rams}},\
  }\href {\doibase 10.1088/1367-2630/aae3ed} {\bibfield  {journal} {\bibinfo
  {journal} {New Journal of Physics}\ }\textbf {\bibinfo {volume} {20}},\
  \bibinfo {pages} {105002} (\bibinfo {year} {2018})}\BibitemShut {NoStop}%
\bibitem [{\citenamefont {Susa}\ \emph
  {et~al.}(2018{\natexlab{a}})\citenamefont {Susa}, \citenamefont {Yamashiro},
  \citenamefont {Yamamoto},\ and\ \citenamefont {Nishimori}}]{Susa2018}%
  \BibitemOpen
  \bibfield  {author} {\bibinfo {author} {\bibfnamefont {Y.}~\bibnamefont
  {Susa}}, \bibinfo {author} {\bibfnamefont {Y.}~\bibnamefont {Yamashiro}},
  \bibinfo {author} {\bibfnamefont {M.}~\bibnamefont {Yamamoto}}, \ and\
  \bibinfo {author} {\bibfnamefont {H.}~\bibnamefont {Nishimori}},\ }\href
  {\doibase 10.7566/JPSJ.87.023002} {\bibfield  {journal} {\bibinfo  {journal}
  {Journal of the Physical Society of Japan}\ }\textbf {\bibinfo {volume}
  {87}},\ \bibinfo {pages} {023002} (\bibinfo {year}
  {2018}{\natexlab{a}})}\BibitemShut {NoStop}%
\bibitem [{\citenamefont {Susa}\ \emph
  {et~al.}(2018{\natexlab{b}})\citenamefont {Susa}, \citenamefont {Yamashiro},
  \citenamefont {Yamamoto}, \citenamefont {Hen}, \citenamefont {Lidar},\ and\
  \citenamefont {Nishimori}}]{Susa2018a}%
  \BibitemOpen
  \bibfield  {author} {\bibinfo {author} {\bibfnamefont {Y.}~\bibnamefont
  {Susa}}, \bibinfo {author} {\bibfnamefont {Y.}~\bibnamefont {Yamashiro}},
  \bibinfo {author} {\bibfnamefont {M.}~\bibnamefont {Yamamoto}}, \bibinfo
  {author} {\bibfnamefont {I.}~\bibnamefont {Hen}}, \bibinfo {author}
  {\bibfnamefont {D.~A.}\ \bibnamefont {Lidar}}, \ and\ \bibinfo {author}
  {\bibfnamefont {H.}~\bibnamefont {Nishimori}},\ }\href {\doibase
  10.1103/PhysRevA.98.042326} {\bibfield  {journal} {\bibinfo  {journal}
  {Physical Review A}\ }\textbf {\bibinfo {volume} {98}},\ \bibinfo {pages}
  {042326} (\bibinfo {year} {2018}{\natexlab{b}})}\BibitemShut {NoStop}%
\bibitem [{\citenamefont {Hartmann}\ and\ \citenamefont
  {Lechner}(2019{\natexlab{a}})}]{Hartmann2019}%
  \BibitemOpen
  \bibfield  {author} {\bibinfo {author} {\bibfnamefont {A.}~\bibnamefont
  {Hartmann}}\ and\ \bibinfo {author} {\bibfnamefont {W.}~\bibnamefont
  {Lechner}},\ }\href {http://arxiv.org/abs/1906.11459} {\  (\bibinfo {year}
  {2019}{\natexlab{a}})},\ \Eprint {http://arxiv.org/abs/1906.11459}
  {arXiv:1906.11459} \BibitemShut {NoStop}%
\bibitem [{\citenamefont {Hatomura}\ and\ \citenamefont
  {Mori}(2018)}]{Hatomura2018b}%
  \BibitemOpen
  \bibfield  {author} {\bibinfo {author} {\bibfnamefont {T.}~\bibnamefont
  {Hatomura}}\ and\ \bibinfo {author} {\bibfnamefont {T.}~\bibnamefont
  {Mori}},\ }\href {\doibase 10.1103/PhysRevE.98.032136} {\bibfield  {journal}
  {\bibinfo  {journal} {Physical Review E}\ }\textbf {\bibinfo {volume} {98}},\
  \bibinfo {pages} {032136} (\bibinfo {year} {2018})}\BibitemShut {NoStop}%
\bibitem [{\citenamefont {Gu{\'{e}}ry-Odelin}\ \emph
  {et~al.}(2019)\citenamefont {Gu{\'{e}}ry-Odelin}, \citenamefont {Ruschhaupt},
  \citenamefont {Kiely}, \citenamefont {Torrontegui}, \citenamefont
  {Mart{\'{i}}nez-Garaot},\ and\ \citenamefont {Muga}}]{Guery-Odelin2019}%
  \BibitemOpen
  \bibfield  {author} {\bibinfo {author} {\bibfnamefont {D.}~\bibnamefont
  {Gu{\'{e}}ry-Odelin}}, \bibinfo {author} {\bibfnamefont {A.}~\bibnamefont
  {Ruschhaupt}}, \bibinfo {author} {\bibfnamefont {A.}~\bibnamefont {Kiely}},
  \bibinfo {author} {\bibfnamefont {E.}~\bibnamefont {Torrontegui}}, \bibinfo
  {author} {\bibfnamefont {S.}~\bibnamefont {Mart{\'{i}}nez-Garaot}}, \ and\
  \bibinfo {author} {\bibfnamefont {J.}~\bibnamefont {Muga}},\ }\href {\doibase
  10.1103/RevModPhys.91.045001} {\bibfield  {journal} {\bibinfo  {journal}
  {Reviews of Modern Physics}\ }\textbf {\bibinfo {volume} {91}},\ \bibinfo
  {pages} {045001} (\bibinfo {year} {2019})}\BibitemShut {NoStop}%
\bibitem [{\citenamefont {Santoro}\ \emph {et~al.}(2002)\citenamefont
  {Santoro}, \citenamefont {Marton{\'{a}}k}, \citenamefont {Tosatti},\ and\
  \citenamefont {Car}}]{Santoro2002}%
  \BibitemOpen
  \bibfield  {author} {\bibinfo {author} {\bibfnamefont {G.~E.}\ \bibnamefont
  {Santoro}}, \bibinfo {author} {\bibfnamefont {R.}~\bibnamefont
  {Marton{\'{a}}k}}, \bibinfo {author} {\bibfnamefont {E.}~\bibnamefont
  {Tosatti}}, \ and\ \bibinfo {author} {\bibfnamefont {R.}~\bibnamefont
  {Car}},\ }\href {\doibase 10.1126/science.1068774} {\bibfield  {journal}
  {\bibinfo  {journal} {Science}\ }\textbf {\bibinfo {volume} {295}},\ \bibinfo
  {pages} {2427} (\bibinfo {year} {2002})}\BibitemShut {NoStop}%
\bibitem [{\citenamefont {Marton{\'{a}}k}\ \emph {et~al.}(2002)\citenamefont
  {Marton{\'{a}}k}, \citenamefont {Santoro},\ and\ \citenamefont
  {Tosatti}}]{Martonak2002}%
  \BibitemOpen
  \bibfield  {author} {\bibinfo {author} {\bibfnamefont {R.}~\bibnamefont
  {Marton{\'{a}}k}}, \bibinfo {author} {\bibfnamefont {G.~E.}\ \bibnamefont
  {Santoro}}, \ and\ \bibinfo {author} {\bibfnamefont {E.}~\bibnamefont
  {Tosatti}},\ }\href {\doibase 10.1103/PhysRevB.66.094203} {\bibfield
  {journal} {\bibinfo  {journal} {Physical Review B}\ }\textbf {\bibinfo
  {volume} {66}},\ \bibinfo {pages} {094203} (\bibinfo {year}
  {2002})}\BibitemShut {NoStop}%
\bibitem [{\citenamefont {del Campo}\ \emph {et~al.}(2012)\citenamefont {del
  Campo}, \citenamefont {Rams},\ and\ \citenamefont {Zurek}}]{DelCampo2012}%
  \BibitemOpen
  \bibfield  {author} {\bibinfo {author} {\bibfnamefont {A.}~\bibnamefont {del
  Campo}}, \bibinfo {author} {\bibfnamefont {M.~M.}\ \bibnamefont {Rams}}, \
  and\ \bibinfo {author} {\bibfnamefont {W.~H.}\ \bibnamefont {Zurek}},\ }\href
  {\doibase 10.1103/PhysRevLett.109.115703} {\bibfield  {journal} {\bibinfo
  {journal} {Physical Review Letters}\ }\textbf {\bibinfo {volume} {109}},\
  \bibinfo {pages} {115703} (\bibinfo {year} {2012})}\BibitemShut {NoStop}%
\bibitem [{\citenamefont {Demirplak}\ and\ \citenamefont
  {Rice}(2003)}]{Demirplak2003}%
  \BibitemOpen
  \bibfield  {author} {\bibinfo {author} {\bibfnamefont {M.}~\bibnamefont
  {Demirplak}}\ and\ \bibinfo {author} {\bibfnamefont {S.~A.}\ \bibnamefont
  {Rice}},\ }\href {\doibase 10.1021/jp030708a} {\bibfield  {journal} {\bibinfo
   {journal} {The Journal of Physical Chemistry A}\ }\textbf {\bibinfo {volume}
  {107}},\ \bibinfo {pages} {9937} (\bibinfo {year} {2003})}\BibitemShut
  {NoStop}%
\bibitem [{\citenamefont {Berry}(2009)}]{Berry2009}%
  \BibitemOpen
  \bibfield  {author} {\bibinfo {author} {\bibfnamefont {M.~V.}\ \bibnamefont
  {Berry}},\ }\href {\doibase 10.1088/1751-8113/42/36/365303} {\bibfield
  {journal} {\bibinfo  {journal} {Journal of Physics A: Mathematical and
  Theoretical}\ }\textbf {\bibinfo {volume} {42}},\ \bibinfo {pages} {365303}
  (\bibinfo {year} {2009})}\BibitemShut {NoStop}%
\bibitem [{\citenamefont {Takahashi}(2017)}]{Takahashi2017}%
  \BibitemOpen
  \bibfield  {author} {\bibinfo {author} {\bibfnamefont {K.}~\bibnamefont
  {Takahashi}},\ }\href {\doibase 10.1103/PhysRevA.95.012309} {\bibfield
  {journal} {\bibinfo  {journal} {Physical Review A}\ }\textbf {\bibinfo
  {volume} {95}},\ \bibinfo {pages} {012309} (\bibinfo {year}
  {2017})}\BibitemShut {NoStop}%
\bibitem [{\citenamefont {Hatomura}(2017)}]{Hatomura2017}%
  \BibitemOpen
  \bibfield  {author} {\bibinfo {author} {\bibfnamefont {T.}~\bibnamefont
  {Hatomura}},\ }\href {\doibase 10.7566/JPSJ.86.094002} {\bibfield  {journal}
  {\bibinfo  {journal} {Journal of the Physical Society of Japan}\ }\textbf
  {\bibinfo {volume} {86}},\ \bibinfo {pages} {094002} (\bibinfo {year}
  {2017})}\BibitemShut {NoStop}%
\bibitem [{\citenamefont {{\"{O}}zg{\"{u}}ler}\ \emph
  {et~al.}(2018)\citenamefont {{\"{O}}zg{\"{u}}ler}, \citenamefont {Joynt},\
  and\ \citenamefont {Vavilov}}]{Ozguler2018}%
  \BibitemOpen
  \bibfield  {author} {\bibinfo {author} {\bibfnamefont {A.~B.}\ \bibnamefont
  {{\"{O}}zg{\"{u}}ler}}, \bibinfo {author} {\bibfnamefont {R.}~\bibnamefont
  {Joynt}}, \ and\ \bibinfo {author} {\bibfnamefont {M.~G.}\ \bibnamefont
  {Vavilov}},\ }\href {\doibase 10.1103/PhysRevA.98.062311} {\bibfield
  {journal} {\bibinfo  {journal} {Physical Review A}\ }\textbf {\bibinfo
  {volume} {98}},\ \bibinfo {pages} {062311} (\bibinfo {year}
  {2018})}\BibitemShut {NoStop}%
\bibitem [{\citenamefont {Hartmann}\ and\ \citenamefont
  {Lechner}(2019{\natexlab{b}})}]{Hartmann2019a}%
  \BibitemOpen
  \bibfield  {author} {\bibinfo {author} {\bibfnamefont {A.}~\bibnamefont
  {Hartmann}}\ and\ \bibinfo {author} {\bibfnamefont {W.}~\bibnamefont
  {Lechner}},\ }\href {\doibase 10.1088/1367-2630/ab14a0} {\bibfield  {journal}
  {\bibinfo  {journal} {New Journal of Physics}\ }\textbf {\bibinfo {volume}
  {21}},\ \bibinfo {pages} {043025} (\bibinfo {year}
  {2019}{\natexlab{b}})}\BibitemShut {NoStop}%
\end{thebibliography}%

\end{document}